\documentclass[A4paper,12pt]{article}
\usepackage[T1]{fontenc}
\usepackage[english]{babel}
\usepackage[cp1250]{inputenc}
\usepackage{epsfig}
\usepackage{amsfonts}
\usepackage{fancyhdr}
\usepackage{syntonly}
\usepackage{graphicx}

\begin{document}
\selectlanguage{english}

\begin{titlepage}
\begin{center}
\vspace*{3cm}

\begin{title}
\bold {\Huge On multiplicity correlations \\
\vspace{0.3cm}in the STAR data}
\end{title}

\vspace{2cm}

\begin{author}
\Large K. FIA{\L}KOWSKI\footnote{e-mail address:
fialkowski@th.if.uj.edu.pl}, R. WIT\footnote{e-mail address:
romuald.wit@uj.edu.pl}

\end{author}

\vspace{1cm}

{\sl M. Smoluchowski Institute of Physics\\ Jagellonian University \\

30-059 Krak{\'o}w, ul.Reymonta 4, Poland}

\vspace{2cm}

\begin{abstract}
The STAR data on the multiplicity correlations between narrow
pseudorapidity bins in the pp and AuAu collisions are discussed. The
PYTHIA 8.145 generator is used for the pp data, and a na\"{i}ve
superposition model is presented for the AuAu data. It is shown that
the PYTHIA generator with default parameter values describes the pp
data reasonably well, whereas the superposition model fails to
reproduce the centrality dependence seen in the data. Some possible
reasons for this failure and a comparison with other models are
presented.

\end{abstract}

\end{center}

\vspace{2cm}


\end{titlepage}

\section{Introduction}

The STAR experiment at the RHIC accelerator has measured the
dependence of the multiplicity correlations between the symmetric
narrow pseudorapidity bins as a function of pseudorapidity distance
for pp and AuAu data at 200 GeV in the nucleon-nucleon
center-of-mass (CM) system \cite{STAR}. The data were taken
separately for different centrality classes in AuAu collisions. The
authors noted that the HIJING model \cite{HIJING} describes
reasonably well the pp data, but fails completely for the central
AuAu data, whereas the predictions of the Parton String Model
\cite{PSM} lie closer to the AuAu data, but fail to describe the
dependence on pseudorapidity distance in the pp data.
\par
The AuAu data were discussed in the subsequent paper by Y.-L. Yan et
al. \cite{Yan}, who compared them with the PACIAE model
\cite{PACIAE} based on the PYTHIA generator \cite{SMS}. The model
overestimates systematically the correlation strength, and the
difference increases strongly with decreasing centrality.
\par
The authors show also the predictions of three other models,
which differ significantly from those of PACIAE model for most
central events, but overestimate the correlations strength for
semiperipheral events as strongly as PACIAE.
\par
In this paper we compare the STAR data and the models mentioned
above with the new version of PYTHIA generator \cite{SMS2} for the
pp data and the na\"{i}ve superposition model for the AuAu data. We
have found that already with the default values of parameters the
PYTHIA 8.145 generator describes reasonably well the pp data. On the
other hand, the superposition model fails completely to describe the
AuAu data. In particular, the centrality dependence of the
correlation strength, which was weaker than in the data in the
PACIAE model, is completely absent in our model.
\par
In the next section we describe the procedures used in the STAR
paper for pp and AuAu data, which subsequently are used by us to
analyze the events generated by PYTHIA and the superposition model.
In Section 3. we show the results, and we discuss them in Section 4.
Some conclusions and perspectives are presented in the last section.

\section{Definitions and procedures}

A standard quantity describing the strength  of correlations between
the phase space bins $1$ and $2$ is the correlation coefficient $b$.

\begin{equation}
\label{eq:b}
 b =  \frac {D^2_{12}}{D_1D_2} = \frac {\overline{n_1 \cdot n_2}-\overline{n_1} \cdot
 \overline{n_2}}{\sqrt{(\overline{n_1^2}-\overline{n_1}^2)(\overline{n_2^2}-\overline{n_2}^2)}},
\end{equation} where $n_1$ and $n_2$ denote the multiplicities in bin $1$ and
$2$, respectively. In the STAR data analysis the pseudorapidity bins
of the width $0.2$ (placed symmetrically in the CM frame) were used.
The distance between the bin centers ranged from $0.2$ to $1.8$.
\par
It is well known that the correlation coefficients in hadron-hadron
scattering increase with energy reflecting mainly the fact that the
multiplicity distributions are much wider than Poissonian and their
widths grow with energy. The STAR authors were mainly interested in
the comparison of the correlation strength for different centrality
classes in heavy ion collisions. These classes were defined by the
range of multiplicity $N$ of charged particles with $p_T > 0.15$
GeV/c in a control bin of unit width, disjunctive with the bins for
which $b$ was measured (to be defined in Section 3).
\par
The $b$ values calculated directly from the formula above reflect
mainly the spread of the average multiplicities within each
centrality class, which is quite large. The ranges of $N$ producing
equal number of events in each class are approximately $>430$,
$320-430$, $230-320$, $155-230$ and $90-155$ for the most central
$0-10\%$, $10-20\%$, $20-30\%$, $30-40\%$ and $40-50\%$ events,
respectively. The relation between $\overline{n_i}$ and $N$ is
approximately given by the ratio of the bin widths: $\overline{n_i}
= 0.2 N$, since $dN/d\eta$ is nearly flat.
\par
Let us consider a distribution of a variable $n$ depending on the
external parameter $N$ in such a way, that $\overline{n}=\alpha N$.
It is well known that in such a case the dispersion may be separated
into two parts: the first one averaging over this parameter, and the
second one reflecting the spread of this parameter. The
corresponding formula reads as follows:
\begin{equation}
\label{eq:D2}
D^2~=~<\overline{n^2}>-<\overline{n}>^2~=~<\overline{n^2}-\overline{n}^2>+<\overline{n}^2>-<\overline{n}>^2~
=~<D^2(N)>+\alpha^2D_N^2 .
\end{equation}
 Here the bar denotes
averaging over the distribution of $n$ for given $N$, $<...>$
denotes averaging over $N$, and in the last term $D_N^2$ denotes the
dispersion of $N$.
\par The authors of STAR performed a more involved procedure to separate
the "dynamical" part of the correlation coefficient from the
dominant "combinatorial" part. In each centrality class and for each
distance between bins they calculated the averages $\overline{n_i}$,
$\overline{n_i^2}$ and $\overline{n_1 \cdot n_2}$ for each value of
$N$ and fitted the dependence on $N$ within each class by the linear
or quadratic formula:
\begin{equation}
\label{eq:fits}
 \overline{n_i}~=~a_i+b_i \cdot
N,~~\overline{n_i^2}~=~c_i+d_i \cdot N+e_i \cdot N^2,~~\overline{n_1
\cdot n_2}~=~c_{12}+d_{12} \cdot N+e_{12} \cdot N^2.
\end{equation}
\par
Using the fitted values of the parameters they calculated these
averages for $N=\overline{N}$ and inserted them into the formula
(\ref{eq:b}). A value of $b$ was obtained for each centrality class
and each value of pseudorapidity bin separation.
\par
One may note that the distributions of $n_i$ for fixed $N$ are
narrow. We assume that
\begin{equation}
\label{eq:narrow} D^2(N)~\sim~\overline{n}(N).
\end{equation}
It is
easy to check that in such a case the procedure described above
gives quite similar results as simple averaging of $b$ over the
distribution of $N$ within each class.
\par
Indeed, let us assume that the formula (\ref{eq:narrow}) holds for
both (identical) distributions of $n_1$ and $n_2$, and a similar
formula holds for $D_{12}^2$. Then the parameters of the fits
(\ref{eq:fits}) obey the following relations:
\begin{equation}
\label{eq:relations}
a_1~=~a_2,~~b_1~=~b_2,~~c_1~=~c_2,~~d_1~=~d_2,~~e_1~=~e_2~=~e_{12}~=~b_1^2.
\end{equation}
Averaging separately $D_{12}^2(N)$ and $D_1^2=D_2^2=D_1D_2$ and
inserting them into formula (\ref{eq:b}) we get in general
\begin{equation}
\label{eq:bav}
b=\frac{<D_{12}^2>}{<D_1^2>}=\frac{(e_{12}-b_1^2)<N^2>+(d_{12}-2a_1b_1)<N>+
c_{12}-a_1^2}{(e_{1}-b_1^2)<N^2>+(d_{1}-2a_1b_1)<N>+ c_{1}-a_1^2},
\end{equation}
whereas from the STAR prescription we obtain
\begin{equation}
\label{eq:b'av}
b'=\frac{D_{12}^2(<N>)}{D_1^2(<N>)}=\frac{(e_{12}-b_1^2)<N>^2+(d_{12}-2a_1b_1)<N>+
c_{12}-a_1^2}{(e_{1}-b_1^2)<N>^2+(d_{1}-2a_1b_1)<N>+ c_{1}-a_1^2}.
\end{equation}
As we see, these two formulae are equivalent when  the conditions
(\ref{eq:relations}) are fulfilled.

\par
For the pp data the STAR Collaboration measured the correlation
coefficient $b$  for each value of bin separation $\Delta \eta$ and
for each value of the multiplicity $N$ in the pseudorapidity range
of two units. Then the result was averaged over the distribution of
$N$. This amounts approximately to using the formula (\ref{eq:bav}).
In fact, all three prescriptions are approximately equivalent.

\section {PYTHIA, superposition model and the data}

\par
To understand the physical meaning of the data presented by the STAR
Collaboration it is useful to perform the same procedures on the
events generated by a Monte Carlo generator. For pp collisions there
are many models describing high energy data. We use here the PYTHIA
8.145 with the default value of parameters. We analyze the generated
events in the same way as described above for the STAR data. We
checked that the PYTHIA results are practically the same when we use
the non-diffractive (ND) and the non-single-diffractive (NSD)
samples of events.

\par
In Fig.1 the STAR data at $E_{CM}=200$GeV and the model predictions
for pp collisions are shown. The HIJING \cite{HIJING} and Parton
String Model (PSM) \cite{PSM} predictions are copied from the STAR
paper \cite{STAR}. The values of $b$ depend rather weakly on the bin
separation $\Delta \eta$ and the HIJING and PYTHIA models describe
them reasonably well, although the $\Delta \eta$ dependence seems to
be too strong in HIJING and too weak in PYTHIA. The PSM model
predicts practically no $\Delta \eta$ dependence, and seems to be
incompatible with the data.

\vspace{0.5cm}

\begin{figure}[h!]
\centerline{ \epsfig{figure=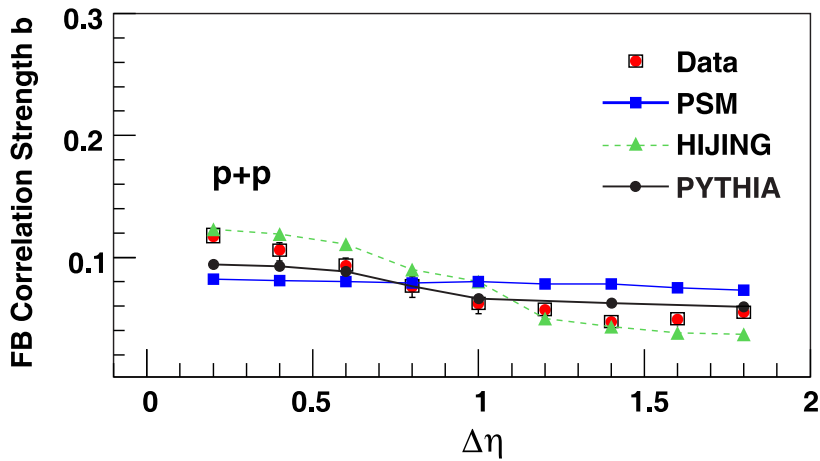,height=6.0cm}}
\caption{\footnotesize \label{pp} The correlation coefficients for
the symmetric pseudorapidity bins of width $0.2$ for the distance
between the bin centers ranging from $0.2$ to $1.8$. Full circles in
open squares, squares, triangles and full circles denote data and
the predictions of PSM, HIJING and PYTHIA models, respectively.
Model predictions are connected by lines to guide the eye.}
\end{figure}

\par
For AuAu collisions the situation is more involved. The centrality
classes were defined by the multiplicity in a control pseudorapidity
bin of the unit length, disjoint with the bins between which the
correlation is measured. For the correlated bins
$(-1,-0.8),~(0.8,1)$ and $(-0.8,-0.6),~(0.6,0.8)$ the control bin is
$(-0.5,0.5)$; for the bins $(-0.2,0),~(0,0.2)$ and
$(-0.4,-0.2),~(0.2,0.4)$ the control bin is made up of two parts:
$(-1,-0.5)$ and $(0.5,1)$. For the bins $(-0.6,-0.4),~(0.4,0.6)$ the
control bin consists of three parts: $(-1,-0.75),~(-0.25,0.25)$ and
$(0.75,1)$. All the generated samples of events were divided into
the equally populated centrality classes, as described in the
previous section.
\par
To generate a sample of AuAu events comparable with the data we used
a na{\"i}ve superposition model. In this model a heavy ion collision
with $N_{part}$ nucleon participants from each nucleus is presented
as a straight superposition of $N_{part}$ pp collisions, each
generated by the PYTHIA. To reproduce a sample of events belonging
to the fixed centrality class we use the known experimental
multiplicity distribution in the control bin $P(N)$. The shape of
the distribution of $N_{part}$ is assumed to be the same as that of
$P(N)$, and the average value $<N_{part}>$ is chosen to reproduce
the value of $<N>$ in this class.
\par
In fact, the shape of $P(N)$ is nearly linear in $N$ for the large
part of the distribution. Only for the most central part of the
distribution ($0-10\%$) it falls down much faster (approximately
Gaussian-like), and for the most peripheral events the maximum is
sharper. This makes the generation of the $P(N_{part})$ quite easy.
In Fig.2 the predictions from such a model for five classes of the
most central events are presented on the right hand side plot. On
the left hand side the STAR data for the same centrality classes are
presented and compared with the results of the PACIAE model
\cite{PACIAE}.

\vspace{-0.7cm}

\begin{figure}[h!]

\centerline{
\epsfig{figure=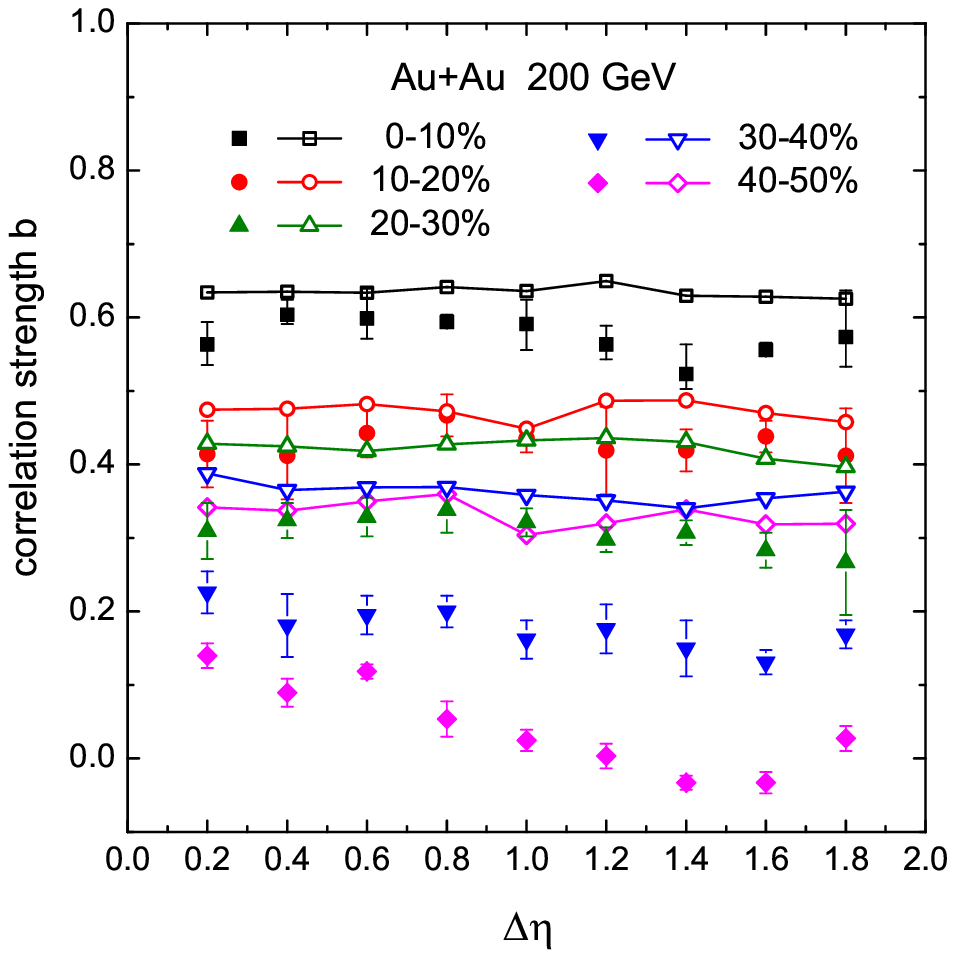,height=9.0cm}
\epsfig{figure=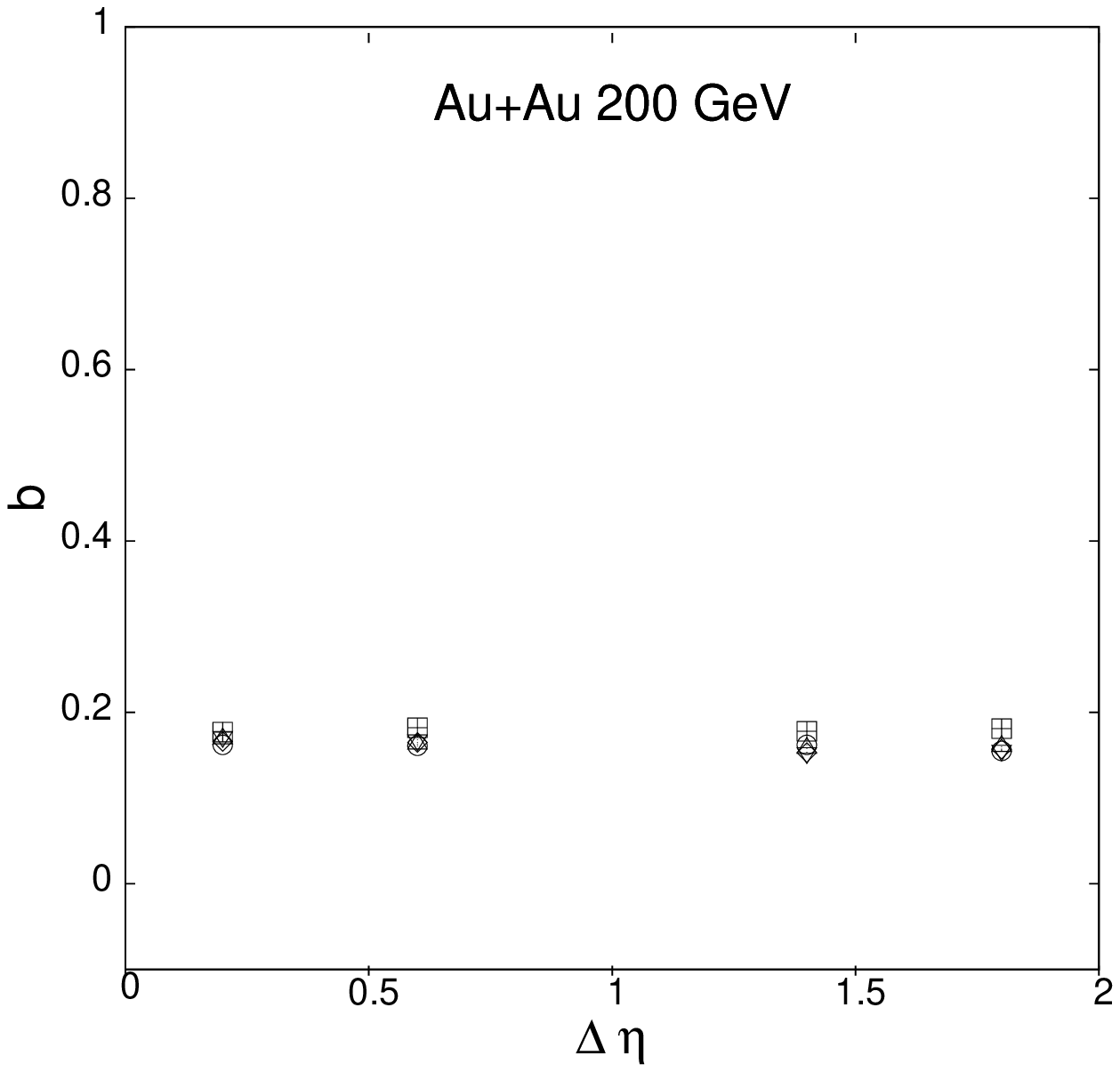,height=7.3cm,width=7.3cm}}

\caption{\footnotesize \label{AuAu} The correlation coefficients $b$
as a function of pseudorapidity distance $\Delta \eta$. On the left:
the STAR data (full symbols) and PACIAE model (open symbols).
Squares, circles, straight and inverted triangles and diamonds are
used for the centrality classes 0-10\%, 10-20\%, 20-30\%, 30-40\%
and 40-50\%, respectively. On the right: our superposition model.
The results for different centrality classes (marked as for the
PACIAE model) are indistinguishable.}
\end{figure}

\par
We see that none of the models describes the data properly. For the
most central class of events the disagreement of the PACIAE model
with data is not very bad, but it grows quickly for the more
peripheral classes. The $\Delta \eta$ dependence of the correlation
coefficient is rather weak and similar in the model and data.
However, the dependence on the centrality is quite strong in the
data and much weaker in the model. Our na{\"i}ve model shows
practically no dependence on centrality. Thus the disagreement with
data is significantly worse. The $\Delta \eta$ dependence is
similarly weak in the data and in the PACIAE model.
\par
One may add that the authors of the PACIAE model compared the data
also with other models \cite{PACIAE}. The spread of their
predictions is quite large and for most central events they bracket
the data. However, the results for most peripheral data considered
(the $40\% - 50\%$ class) for all the models are much higher than
the data, which exhibit in fact the negative values of the
correlation coefficient $b$ for the most distant bins.

\section {Discussion}

We have seen that the pp data analyzed separately for the given
multiplicity $N$ in the available pseudorapidity range and then
averaged over the distribution of $N$ are reasonably well described
by different models. Both the value of the correlation coefficient
$b \approx 0.1$ and the weak dependence on the psudorapidity
separation $\Delta \eta$ appear without any tuning.
\par
We should remember that the procedure used by STAR was designed to
remove the influence of the global multiplicity fluctuations from
the correlation effects. In fact, for the standard inclusive
definition of the correlation coefficient $b$ the dispersions in the
numerator and denominator of formula (\ref{eq:b}) are dominated by
these fluctuations. The value of $b$ is then twice bigger than $0.1$
and increases with the bin width. Thus a valid question is: what are
the correlations measured by the STAR procedure?
\par
To clarify this point, let us consider an oversimplified model where
there are no correlations for a fixed value of $N$. Then the
distribution of the multiplicity $n$ inside a narrow pseudorapidity
bin of the width $y$ is given by the binomial distribution
\begin{equation}
\label{eq:bin} P(n) ={N \choose n} p^n(1-p)^{N-n};~p = ~<n>/N.
\end{equation}
 The generating function of this distribution is
\begin{equation}
\label{eq:gf1} g(z) \equiv \sum_n z^n P(n) =(pz+1-p)^N.
\end{equation}
Remember that the factorial moments of the distribution
(\ref{eq:bin}) are given by
\begin{equation}
\label{eq:mom1} <n_q>~ \equiv \frac{n!}{(n-q)!} =\frac {dg(z)}{dz^q}
\Bigg\vert_{z=1}=p^qN_q
\end{equation}
It is easy to generalize these formulae for the case of two bins.
The corresponding generating function is then
\begin{equation}
\label{eq:gf2} g(z_1,z_2) \equiv \sum_{n_1,n_2}
z_1^{n_1}z_2^{n_2}P(n_1,n_2)=[p(z_1+z_2)+1-2p]^N.
\end{equation}
The factorial moments for both bins can be calculated as in equation
(\ref{eq:mom1}) and the average product of multiplicities as
\begin{equation}
\label{eq:product} <n_1n_2>~ \equiv \frac
{d^2g(z_1,z_2)}{dz_1dz_2}\Bigg\vert_{z_1=1,z_2=1}=p^2N(N-1).
\end{equation}
This results in the simple formulae for the dispersions
\begin{equation}
\label{eq:disp} D_1^2=D_2^2=p(1-p)N,~~D_{12}^2=-p^2N
\end{equation}
and for the correlation coefficient
\begin{equation}
\label{eq:bbin} b =\frac{D_{12}^2}{D_1^2}=\frac{-p}{1-p}.
\end{equation}
Since this formula does not depend on $N$, the averaging of $b$ over
the distribution of $N$ will give the same result. This will not
change if we average the numerator and denominator in
(\ref{eq:bbin}) separately.
\par
We see that this simplified model disagrees strongly with data: the
values of $b$ are always negative, whereas in the data they are
positive. Possible effects which introduce the positive correlations
are:

\begin{itemize}
\item the existence of resonances, probably responsible for the
dependence on $\Delta \eta$
\item the existence of jets and minijets, possible dijet
correlations
\item the contribution of diffractive events (important for small
$N$, decreasing for large $N$)
\item Bose-Einstein interference, which may be important for adjacent bins.
\end{itemize}

Therefore we may conclude that the STAR procedure allows to measure
the influence of these effects. They are significant, as they
reverse the sign of the correlation coefficient. As indicated above,
the realistic MC models describe them quite well.

\par
For the AuAu collisions none of the considered models describes the
data properly. The centrality dependence of the correlation
coefficient $b$ seen in the data is significantly stronger than in
the models. Moreover, our na{\"i}ve model predicts practically no
centrality dependence at all and thus the disagreement with the data
is even worse than for other models.
\par
Therefore one should ask what is the difference between our model
and the other models. In our model it is assumed explicitly that the
number of participating nucleons is the same in both colliding
nuclei. It seems that this assumption suppresses the centrality
dependence of the correlation coefficient.
\par
This observation suggests that the data of STAR Collaboration, when
interpreted within the superposition picture, measure mainly the
fluctuations in the number of participating nucleons $N_{part}$ in
both nuclei. For the most central class of events these fluctuations
seem to be strongly correlated. With decreasing centrality this
correlation becomes weaker both in the data and in the models.
\par
One can understand this effect qualitatively in the framework of a
"hidden asymmetry" picture \cite{BZ}, in which the dominant
configuration is strongly asymmetric. There are few events in which
the number of participants in both colliding nuclei is the same or
similar. The correlation coefficient $b$ is positive and large for
most central events, where the number of participants in both nuclei
is nearly maximal (and thus approximately the same). The centrality
is defined by the multiplicity in a symmetric control bin close to
$\eta=0$. When this multiplicity decreases, the asymmetric
configurations start to dominate, which results in decreasing values
of $b$.
\begin{figure}[h!]
\centerline{ \epsfig{figure=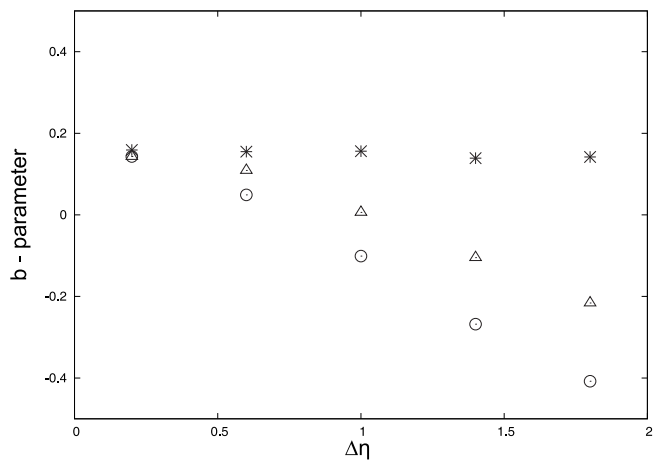,height=6.0cm}}
\caption{\footnotesize \label{AuAu} The correlation coefficients $b$
as a function of pseudorapidity distance $\Delta \eta$  for the
$40-50\%$ class. Asterisks, triangles and circles are for the
original model, triangular and flat distributions of $N_F$,
respectively.}
\end{figure}
This is illustrated in Fig.3, where our toy model predictions for
the $40-50\%$ class are compared with its two modified versions. In
these versions the particles from each $p-p$ event generated by
PYTHIA are divided randomly into the fragmentation products of two
"wounded" nucleons. One  of such "half-events"  populates mainly the
forward CM hemisphere and the other one backward one. The AuAu event
is constructed from the (generally unequal) numbers of such
half-events $N_F$ and $N_B$. For a given sum of these numbers we
tried the distribution of $N_F$ to be flat or to have a triangular
shape.

\par
We see that it is quite easy to explain the decrease of $b$ with the
pseudorapidity distance, even to the values much more negative than
in the data. However, our model fails to explain the strong increase
of $b$ for most central events and the adjacent bins.
\par
This dependence may be reproduced qualitatively in PACIAE and other
similar models, because they are not "pure" superposition models.
The partons from initial nucleons interact forming the parton
cascades which then hadronize. With increasing centrality the number
of parton cascades increases. Apparently, the corresponding increase
of the numerator in formula (\ref{eq:b}) is faster than the
corresponding increase of the denominator.
\par
However, the decrease for non-central events is underestimated in
all the models. The negative values of $b$ in the data for the
$40-50\%$ class, which correspond to the fact that for this class of
events the positive fluctuation of $N_{part}$ in one nucleus is
correlated with a negative fluctuation in the second nucleus, are
not reproduced. This conforms with the idea of "hidden asymmetry".
If asymmetric configurations are prevalent for semiperipheral
events, the anticorrelation is natural.

\section {Conclusions and outlook}
\par
The failure of superposition models is usually explained by the
collective effects. An example of such an effect is the quark-gluon
plasma formation, which should be most visible in the central
events. However, the data for the most central $10\%$ of events are
quite well described by the models where no plasma is formed
\cite{Yan}. As noted above, the biggest discrepancy is observed for
semi-peripheral events. Other collective effects, as the
hydrodynamical flow in the hadronic phase, are strongest for this
class of events. On the other hand, it is not clear how such effects
may be responsible for the discrepancy between the models and data.
It seems more likely that the explanation is related to the "hidden
asymmetry" of the interacting nucleons.
\par
The negative correlation for the semiperipheral collisions are not
the only unexpected feature of the STAR data. It was noted
\cite{Bzdak} that the large values of $b$ for the most central class
of events suggest that the correlation between the distant bins "B"
and "F" is stronger than the correlations between these bins and the
central control bin. This contradicts the generally accepted feature
of the uniform decrease of correlations with increasing distance
between bins. Therefore we conclude that the centrality dependence
of $b$ reported by STAR seems to be surprisingly strong. It would be
interesting to measure this dependence collecting further data.

\section {Acknowledgements}

We are grateful to Andrzej Bia{\l}as and Barbara Wosiek for helpful
remarks.

\end{document}